\documentclass[11pt,preprint]{article}
\usepackage[dvips]{graphicx}
\usepackage{setspace}
\def\farcs{\hbox{$.\!\!^{\prime\prime}$}}
\def\la{\mathrel{\mathchoice {\vcenter{\offinterlineskip\halign{\hfil
$\displaystyle##$\hfil\cr<\cr\sim\cr}}}
{\vcenter{\offinterlineskip\halign{\hfil$\textstyle##$\hfil\cr
<\cr\sim\cr}}}
{\vcenter{\offinterlineskip\halign{\hfil$\scriptstyle##$\hfil\cr
<\cr\sim\cr}}}
{\vcenter{\offinterlineskip\halign{\hfil$\scriptscriptstyle##$\hfil\cr
<\cr\sim\cr}}}}}
\def\ga{\mathrel{\mathchoice {\vcenter{\offinterlineskip\halign{\hfil
$\displaystyle##$\hfil\cr>\cr\sim\cr}}}
{\vcenter{\offinterlineskip\halign{\hfil$\textstyle##$\hfil\cr
>\cr\sim\cr}}}
{\vcenter{\offinterlineskip\halign{\hfil$\scriptstyle##$\hfil\cr
>\cr\sim\cr}}}
{\vcenter{\offinterlineskip\halign{\hfil$\scriptscriptstyle##$\hfil\cr
>\cr\sim\cr}}}}}

\def\Ncol{\mbox{$N_{\rm H_2}$}}
\def\vinf{\mbox{$v_{\rm inf}$}}
\def\Macc{\mbox{$\dot{M}_{\rm acc}$}}
\def\Minh{\mbox{$\dot{M}_{\rm inh}$}}
\def\Mhii{\mbox{$\dot{M}_{\rm HII}$}}

\def\kms{\mbox{km~s$^{-1}$}}
\def\AMM{NH$_3$}
\def\MCN{\mbox{CH$_3$CN}}
\def\HII{H{\sc ii}}

\newcommand{\lesssim}{\mathrel{\hbox{\rlap{\hbox{\lower4pt\hbox{$\sim$}}}\hbox{$<$}}}}

\begin{document}

\title{\huge {\bf Infall of gas as the formation mechanism of stars up to 20
times more massive than the Sun}}

\author{Maria T.~Beltr\'an$^1$, 
Riccardo Cesaroni$^2$, Claudio Codella$^3$, \\
Leonardo Testi$^2$, Ray S.~Furuya$^4$ ~\& Luca Olmi$^3$}  

\maketitle

$^1$ Departament d'Astronomia i Meteorologia, 
Universitat de Barcelona, Av.~Diagonal, 647, 08028, Barcelona, Catalunya, Spain 

$^2$  Osservatorio Astrofisico di Arcetri, Largo E. Fermi 5, 50125 Firenze,
Italy 

$^3$ Istituto di Radioastronomia, INAF, Sezione di Firenze, Largo E. Fermi 5, 50125 Firenze,
Italy 

$^4$  Subaru Telescope, National Astronomical Observatory
of Japan, 650 North Aohoku Place, Hilo, HI 96720, USA \\


%

{\bf Theory predicts and observations confirm that low-mass stars (like the Sun)
in their early life grow by accreting gas from the surrounding material. But for
stars $\sim 10$ times more massive than the Sun ($\sim 10~M_\odot$), the
powerful stellar radiation is expected to inhibit accretion$^1$ and thus limit the
growth of their mass. Clearly, stars with masses $>10~M_\odot$ exist, so there
must be a way for them to form. The problem may be solved by
non-spherical accretion$^{2,3}$, which allows some of the stellar photons to
escape along the symmetry axis where the density is lower.  The recent
detection of rotating disks$^{4-6}$ and toroids$^7$ around very young massive
stars has lent support to the idea that high-mass ($\ga8~M_\odot$) stars
could form in this way. Here we report observations of an ammonia line
towards a high-mass star forming region. We conclude from the data that the gas
is falling inwards towards a very young star of $\sim20~M_\odot$, in line
with theoretical predictions of non-spherical accretion.}

In the work we report here, we have revealed the simultaneous presence of three
elements in the same massive object: outflow, rotation, and infall. Although
evidence of circumstellar disks  or toroids in massive objects had been provided
by a number of observational studies$^{4-10}$ (see Ref.~11 for a review), it
remained to be proved that these are accretion structures, sustaining the growth
of the central star(s). Evidence of infall had been found only in a very limited
number of massive objects$^{12-17}$, but in no case had the simultaneous
presence of rotation, outflow, and infall towards an O-B (proto)star been
established. Our findings hence represent a substantial advance in this field.

The source under investigation is G24.78+0.08, a massive star forming region
located at a distance of 7.7~kpc, studied by us in great detail$^{7,18-21}$.
This contains a cluster of Young Stellar Objects (YSOs), three of them
associated with rotating toroids. In one of these, G24~A1, the presence of an
early-type star is witnessed by a hypercompact \HII\ region (M.T.B. {\it et
al.}, manuscript in preparation) with a diameter $\la$$0\farcs2$ ($\la$1,500~AU)
and located at the  geometrical centre of the toroid (see Fig.~1). The continuum
spectrum resembles that of a classical (Str\"omgren) \HII\ region around a
zero-age main sequence star of spectral type O9.5$^{18}$, corresponding to a
mass of $\sim 20~M_\odot$, a  luminosity of $\sim$33,000~$L_\odot$ and a Lyman
continuum of $5\times10^{47}$~sec$^{-1}$. Being the \HII\ region very bright at
centimeter wavelengths ($\ga$2,000~K), it is easy to observe the colder
molecular gas ($\sim$100~K) in absorption against it. If the toroid is not only
rotating, but also accreting onto the central star, one expects to see
absorption at positive velocities (i.e.\ red-shifted by Doppler effect) relative
to the stellar velocity$^{12-15}$. The latter ($\sim$110.8~\kms) is equal to the
mean velocity of the gas in the toroid, estimated from molecules whose lines are
not affected by absorption$^{21}$, such as methyl cyanide (\MCN).

With this in mind, we observed simultaneously the continuum emission and  
the ammonia (\AMM) (2,2) inversion transition at  1.3-cm wavelength
using the Very Large Array (VLA) interferometer of
the National Radio Astronomy Observatory (NRAO) in the B configuration. As
expected, the line is seen in absorption towards the continuum of the
\HII\ region: this is shown by the white contours in Fig.~1a,
where one can also see the emission (black contours) detected towards the
adjacent toroid G24~A2, not discussed here.  Clearly, the peak of
absorption as well as the peak of the 1.3-cm continuum emission (i.e.\ of the
\HII\ region) lie at the center of the toroid previously imaged by us (yellow
line) and along the axis of the associated bipolar outflow (black arrows).
The velocity gradient in the rotating toroid is shown in
Fig.~1b, overlaied on an image and contour map of the \MCN\ emission at 1.4-mm 
wavelength previously obtained by us$^{7,21}$ with the IRAM Plateau de Bure
interferometer (PdBI) in the most extended configuration.

Figure~2 illustrates the most important finding of our study. In order to
understand this figure, one must keep in mind that the \AMM\ inversion
transitions have a complex structure, made of a ``main'' component and four
``hyperfine satellites''. The former line is intrinsically stronger than the latter
ones by a factor $\sim$14, so that, roughly speaking, emission in the main line is representative
of lower gas densities, whereas the satellites trace the densest material.
In Fig.~2 we show position-velocity plots for both the main line (Fig.~2a)
and the mean satellite emission (Fig.~2b), where the offset in position
is measured
along the plane of the rotating toroid. In Fig.~2c we show the
same plot for the \MCN(12--11) $K=3$ line which was used by us to trace
rotation in the toroid$^{7}$.
Three considerations are in order: 1) Towards the hypercompact \HII\ region,
the satellite absorption is strongly biased towards positive velocities,
i.e. red-shifted with respect to the velocity of the star; 2) The \AMM\ main
line is seen in absorption at both positive and negative velocities, but the
blue-shifted absorption is fainter, broader, and more optically thin than the
red-shifted one; 3) The velocity gradient seen in the \MCN\ line (outlined
by the green line in Fig.~2c), is detected also in the \AMM\ main line (green
line across the two emission peaks in Fig.~2a), thus confirming the presence
of rotation.

The important conclusion is that the densest gas seen in the satellite
absorption is moving away from the observer at a speed of
$\sim$2~\kms\ (see Fig.~2b) towards the hypercompact \HII\ region,
namely towards the
star. This fact proves the presence of infall onto the O-type star at the
center of the toroid. On the other hand, the faint and broad blue-shifted absorption
seen in the main line is very likely due to the low-column density gas in the lobe
of the molecular outflow that is moving towards the observer: this is
indicated by the lower main line/satellite ratio (i.e. lower optical depth)
observed in the blue line wing with respect to the one in the red-shifted
absorption.
Incidentally, we note that the broad blue-shifted absorption can be used
to discriminate between sources G24~A1 and G24~A2 as origin of the outflow.
Since the blue-shifted lobe is directed towards NW$^{19}$, it cannot originate
from G24~A2 or it would not pass in front of G24~A1. We hence believe that
the star in G24~A1 is powering the flow.

What are the implications of the detection of infall in G24~A1?  Using the
properties of the \AMM(2,2) transition and an estimate of the gas
temperature$^{21}$, one can derive the column density of the infalling gas,
\Ncol, from the ratio between the main line intensity and that of the
satellites$^{22}$, assuming an \AMM\ abundance relative to H$_2$ of
$10^{-6}$--$10^{-7}$ (refs 18, 23). From \Ncol\ and the infall velocity,
$\vinf\simeq2$~\kms, assuming free-fall one obtains the mass accretion rate
onto the star inside a solid angle $\Omega$:
$\Macc\simeq\Omega/(4\,\pi)\,(4\times10^{-4}$--$10^{-2})~M_\odot\,$yr$^{-1}$.
A detailed derivation of the mass accretion rate is presented in Supplementary
Information. The range of values reflects the uncertainty on \Ncol\ and on the radius at
which \vinf\ is measured. Noticeably, \Macc\ is much larger than the critical
rate above which formation of an \HII\ region is inhibited$^{24}$ if
accretion is spherically symmetric (i.e. for $\Omega=4\,\pi$). For an O9.5
star this is $\Minh\simeq8\times10^{-6}~M_\odot\,$yr$^{-1}$, much less than
\Macc: this should suffice to prevent the formation of an \HII\ region. The fact
that, instead, an \HII\ region is detected can be explained only if
the accretion is not spherically symmetric ($\Omega<4\,\pi$), which in
turn supports the existence of a circumstellar disk-like structure, detected
by us on a larger scale as a toroid.

It is also interesting to estimate the rate needed for the ram pressure to
overcome the thermal pressure of the \HII\ region
($\Mhii\simeq\Omega/(4\,\pi)\,2\times10^{-4}~M_\odot\,$yr$^{-1}$). This has
been obtained assuming an electron density and diameter of the \HII\ region
respectively of $\sim10^6$~cm$^{-3}$ $^{(18)}$ and $0\farcs2$ (see above) or
0.0075~pc.  One sees that $\Macc\ge\Mhii$ independently of $\Omega$, which
indicates that the accretion is large enough to brake or slow down the
expansion of the \HII\ region through the toroid. This conclusion is
strengthened if one takes into account the effect of stellar gravity, likely
non-negligible in such a small \HII\ region. Detailed models support this
conclusion, and predict that hypercompact \HII\ regions could be long lived
because of trapping of the ionized gas$^{25}$ owing to infall in the
gravitational field of the star.

We now consider the effect of our findings on current theories of massive star
formation. Stars as massive as $\sim8~M_\odot$ are believed to reach the
zero-age main sequence still accreting.  At this stage, their powerful
radiation pressure appears to be strong enough to halt the infalling
material$^{1,26}$, which should inhibit further growth of the star beyond the
limit of $\sim8~M_\odot$.  In order to solve this problem merging of low-mass
stars has been proposed$^{27}$, although -- even in the most favourable case of
induced binary mergers -- to be effective this scenario requires  very large
stellar densities, at least of the order of $10^6$~stars~pc$^{-3}$.
Increasing the mass accretion rate may be a viable theoretical solution to
overcome the radiation pressure with the ram pressure of the infalling
material.  Moreover, this would solve the lifetime problem$^{28}$ related to
the fact that with accretion rates typical of low-mass stars
($10^{-5}~M_\odot$~yr$^{-1}$) the time needed to form a star $>10~M_\odot$
would be too long ($>10^6$~yr), comparable to the entire lifetime of the
star.  Higher rates can be achieved either by assuming a large turbulent
pressure$^{28,29}$ in the molecular cores forming O-B type stars, or
deepening the gravitational well$^{30}$ around a massive star by surrounding
it with a tight cluster of lower mass companions. Non-spherical accretion may
also soften the problem$^{2,3}$: focusing accretion into a circumstellar disk
reduces significantly the radiation pressure force along the equator by
beaming the radiation in the direction of the poles.

In our case, direct accretion onto the central massive (proto-)star has not
yet been detected, so that the system may be in a (relatively) late phase,
when the final mass of the central object has been assembled and the
infalling material is being deflected in the outflow instead of being
accreted. Nevertheless, our detection of infall towards G24~A1 is an
important milestone towards an observational constraint of the theoretical
models. In fact, even if the very accretion phase were finished, the
presence of a rotating toroid infalling on a hypercompact \HII\ region,
located at the base of a powerful collimated bipolar flow, is
broadly in agreement with the expectations of the non-spherical accretion
scenario for the formation of massive stars. Our results thus suggest that
this is a plausible mechanism for the formation of stars at least as massive
as 20~$M_\odot$. More `exotic' mechanisms (such as stellar mergers) might
still be required to form stars of even higher mass.

\vspace{1cm}

1. Kahn, F.~D.\ Cocoons around early-type stars.\  {\em Astron.~Astrophys.} {\bf 37}, 
149-162 (1974). 

2. Yorke, H.~W. \& Sonnhalter, C.\ On the formation of massive 
stars.\  {\em Astrophys.~J.} {\bf 569}, 846-862 (2002). 

3. Krumholz, M.~R., McKee, C.~F., \& Klein, R.~I.\ How 
protostellar outflows help massive stars form.\  {\em Astrophys.~J.} {\bf 618}, 
L33-L36 (2005). 

4. Cesaroni, R. {\em et al.} A study of the Keplerian accretion disk and precessing outflow in the 
massive protostar IRAS 20126+4104.\  {\em Astron.~Astrophys.} {\bf 434}, 1039-1054 
(2005). 

5. Chini, R. {\em et al.} The formation of a massive protostar through the disk
accretion of gas.\ {\em Nature} {\bf 429}, 155-157 (2004).

6. Patel, N. {\em et al.} A disk of dust and molecular gas around a high-mass
protostar.\ {\em Nature} {\bf 437}, 109-111 (2005). 

7. Beltr{\'a}n, M.~T. {\em et al.}  Rotating disks in high-mass young stellar 
objects.\  {\em Astrophys.~J.} {\bf 601}, L187-L190 (2004). 


8. Cesaroni, R., Felli, M., Testi, L., Walmsley, C.~M. \& Olmi, L.\ The
disk-outflow system around the high-mass (proto)star IRAS 20126+4104.\ 
{\em Astron.~Astrophys.}  {\bf 325}, 725-744 (1997).


9. Jiang, Z.  {\em et al.} A circumstellar disk associated with a massive
protostellar object.\ {\em Nature} {\bf 437}, 112-115 (2005).

10. Cesaroni, R.\ in {\em Massive Star Birth: A Crossroads of Astrophysics} (eds Cesaroni,
R., Felli, M., Churchwell, E. \& Walmsley, M.), 59-69 (Proc. {\em IAU Symp.} {\bf 227}, Univ.\
Press, Cambridge 2005). 

11. Cesaroni, R.\ Outflow, infall, and rotation in high-mass star forming 
regions.\  {\em Ap\&SS} {\bf 295}, 5-17 (2005). 

12. Ho, P.~T.~P. \& Haschick, A.~D.\ Molecular clouds associated with compact 
\HII\  regions. III. Spin-up and collapse in the core of G10.6$-$0.4. {\em
Astrophys.~J.}  {\bf 304}, 501-514 (1986).

13. Keto, E.~R., Ho, P.~T.~P. \& Haschick, A.~D.\ Temperature and density
structure of the collapsing core of G10.6$-$0.4. {\em
Astrophys.~J.}  {\bf 318}, 712-728 (1987).

14. Zhang, Q. \& Ho, P.~T.~P.\ Dynamical collapse in W51 massive cores: 
NH$_3$ observations.\  {\em Astrophys.~J.} {\bf 488}, 241-257 (1997). 

15. Hofner, P., Peterson, S. \& Cesaroni, R.\ Ammonia absorption 
toward the ultracompact \HII\ regions G45.12+0.13 and G45.47+0.05.\  {\em 
Astrophys.~J.} {\bf 514}, 899-908 (1999). 

16. Sollins, P. \& Ho, P.\ T.\ P.\ The molecular accretion flow in G10.6$-$0.4.
{\em Astrophys.~J.}  {\bf 630}, 987-995 (2005). 

17. Zhang, Q.\ in {\em Massive Star Birth: A Crossroads of Astrophysics} (eds Cesaroni, R.,
Felli, M., Churchwell, E. \& Walmsley, M.), 135-144 (Proc. {\em IAU Symp.} {\bf 227}, Univ.\ Press, Cambridge 2005).


18. Codella, C., Testi, L. \& Cesaroni, R.\ The molecular environment of 
H$_2$O masers: VLA ammonia observations.\  {\em Astron.~Astrophys.} {\bf 325}, 282-294 
(1997). 
 
19. Furuya, R.~S. {\em et al.} G24.78+0.08: A cluster of high-mass (proto)stars.\  {\em Astron.~Astrophys.} 
{\bf 390}, L1-L4 (2002). 
 
20. Cesaroni, R., Codella, C., Furuya, R.~S. \& Testi, L.\ Anatomy of a 
high-mass star forming cloud: The G24.78+0.08 (proto)stellar cluster.\  
{\em Astron.~Astrophys.} {\bf 401}, 227-242 (2003). 
 
21. Beltr{\'a}n, M.~T. {\em et al.}  A detailed study of the rotating toroids in 
G31.41+0.31 and G24.78+0.08.\  {\em Astron.~Astrophys.} {\bf 435}, 901-925 (2005). 
 
22. Ungerechts, H., Winnewisser, G. \& Walmsley, C.~M.\ Ammonia observations 
and temperatures in the S140/L1204 molecular cloud.\  {\em Astron.~Astrophys.} {\bf 157}, 
207-216 (1986). 
 
23. Cesaroni, R., Churchwell, E., Hofner, P., Walmsley, C.~M. \& Kurtz, S.\ 
Hot ammonia towards compact \HII\ regions.\  {\em Astron.~Astrophys.} {\bf 288}, 903-920 
(1994). 
 
24. Walmsley, M.\ Dense cores in molecular clouds ({\em Rev.\ Mexicana Astron.\
Astrofis., Conf.\ Ser.}, Vol.\ {\bf 1}, 137--148 (1995). 

25. Keto, E.\ On the evolution of ultracompact \HII\ regions.\  {\em Astrophys.~J.} {\bf 
580}, 980-986 (2002).

26. Wolfire, M.~G. \& Cassinelli, J.~P.\ Conditions for the formation of 
massive stars.\  {\em Astrophys.~J.} {\bf 319}, 850-867 (1987).

27. Bonnell, I.~A. \& Bate, M.~R.\ Binary systems and stellar mergers in 
massive star formation.\  {\em Mon.~Not.~R.~Astron.~Soc.} {\bf 362}, 915-920
(2005). 

28. McKee, C.~F. \& Tan, J.~C.\  The Formation of Massive Stars from Turbulent
Cores. {\em Astrophys.~J.} {\bf 585}, 850-871 (2003). 

29. McKee, C.~F. \& Tan, J.~C.\ Massive star formation in 
100,000 years from turbulent and pressurized molecular clouds.\  {\em 
Nature} {\bf 416}, 59-61 (2002). 

30. Bonnell, I.~A., Vine, S.~G. \& Bate, M.~R.\ Massive star 
formation: nurture, not nature.\  {\em Mon.~Not.~R.~Astron.~Soc.} {\bf 349}, 
735-741 (2004).

\vspace{1cm}

{\bf Supplementary Information} is linked to the online version of the paper at www.nature.com/nature.

{\bf Acknowlegments} NRAO is operated by
Associated University, Inc., under contract with the National Science
Foundation. It is a pleasure to thank Dr. Paul Ho and another anonymous referee
for their constructive criticisms which greatly improved the presentation of our
results.

{\bf Correspondence} and requests for materials should be addressed to M.T.B.
(mbeltran@am.ub.es).

\clearpage

Figure 1: {\bf Absorption and emission by molecular gas towards the hypercompact \HII\
 region G24~A1.}
 {\bf a}, Map of the intensity integrated under the \AMM(2,2) inversion
 line (colour image). Black contours indicate positive intensity (emission),
 while white contours are negative (absorption). The white filled circle
 marks the position of the hypercompact \HII\ region detected in the continuum
 emission at
 1.3-cm wavelength$^{17}$ (Beltr\'an et al. in preparation).
 The black arrows outline the direction
 of the bipolar outflow, ejected along the axis of the disk. The disk
 plane is denoted by the yellow line. Note the deep absorption
 towards the \HII\ region, at the center of the toroid.
  The final maps have been reconstructed with a circular synthesized beam of
$0\farcs8 \times0\farcs8$, which is represented by the hatched circle.  
 The  1-$\sigma$ noise of the integrated
\AMM(2,2) line is 0.6~mJy\,beam$^{-1}$.
 Contour levels start at a $\pm$3-$\sigma$ level and vary by 3-$\sigma$
 (positive contours) and 12-$\sigma$ (negative contours). 
 {\bf b}, Map (image and contours) of the intensity integrated under the
 \MCN(12--11) $K=3$ line$^{21}$
 outlining the toroids in
 G24~A1 and G24~A2 (the latter is not discussed in this article). The colour
 scale corresponds to the velocity field of the toroid in G24~A1: note how the
 velocity changes gradually from red-shifted to blue-shifted, consistently
 with rotation about the outflow axis.
 A velocity gradient, similar to that in G24~A1,
 has also been detected$^7$ in the other core G24~A2, but is neither shown nor discussed here.
 Symbols have the same meaning as above.
 The resulting synthesized beam is $1\farcs2 \times0\farcs5$ at 
 PA = $-174^{\circ}$ and is represented by the hatched ellipse at the bottom
 right. The  1-$\sigma$ noise
of the intensity integrated under the \MCN(12--11) $K=3$ line is 35
mJy\,beam$^{-1}$.
 Contour levels start at 0.25 Jy\,beam$^{-1}$ and increase by 0.2
 Jy\,beam$^{-1}$ . $\delta$, declination; $\alpha$, right ascension.
 
\vspace{1cm}

Figure 2: {\bf Velocity field in the massive toroid G24~A1.}
 {\bf a}, Position-velocity plot of the \AMM(2,2) main line intensity. The
 positional cut is made along the plane of the disk, PA=$-135^\circ$, namely along the
 yellow line in Fig.~1. Both offset and velocity are computed with respect to
 the star. 
 Black and white contours correspond respectively to
 emission and absorption.  The
 continuum has been subtracted from the line emission.
 The  1-$\sigma$ noise is 1~mJy\,beam$^{-1}$.
 Contour levels start at a 3-$\sigma$ level and increase by 3-$\sigma$
 (positive contours) and 6-$\sigma$ (negative contours). 
 The error bars
 denote the angular and spectral resolution.
 The original spectral resolution was $\sim0.31$~\kms,
which was degraded by a factor 2 to improve the S/N by
averaging pairs of adjacent channels.
 Note the velocity gradient
 outlined by the green line.
 {\bf b}, As {\bf a} but for the \AMM(2,2) satellite line. Note how the
 absorption peak is red-shifted with respect to the stellar velocity.
 {\bf c}, As {\bf a} but for the \MCN(12-11) $K=3$ line: the velocity
 gradient outlined by the green line is indicative of rotation$^{21}$.
 The  1-$\sigma$ noise is 48 mJy\,beam$^{-1}$. 
\clearpage

\begin{figure}
\centerline{\resizebox{8.9cm}{!}{\rotatebox{0}{\includegraphics{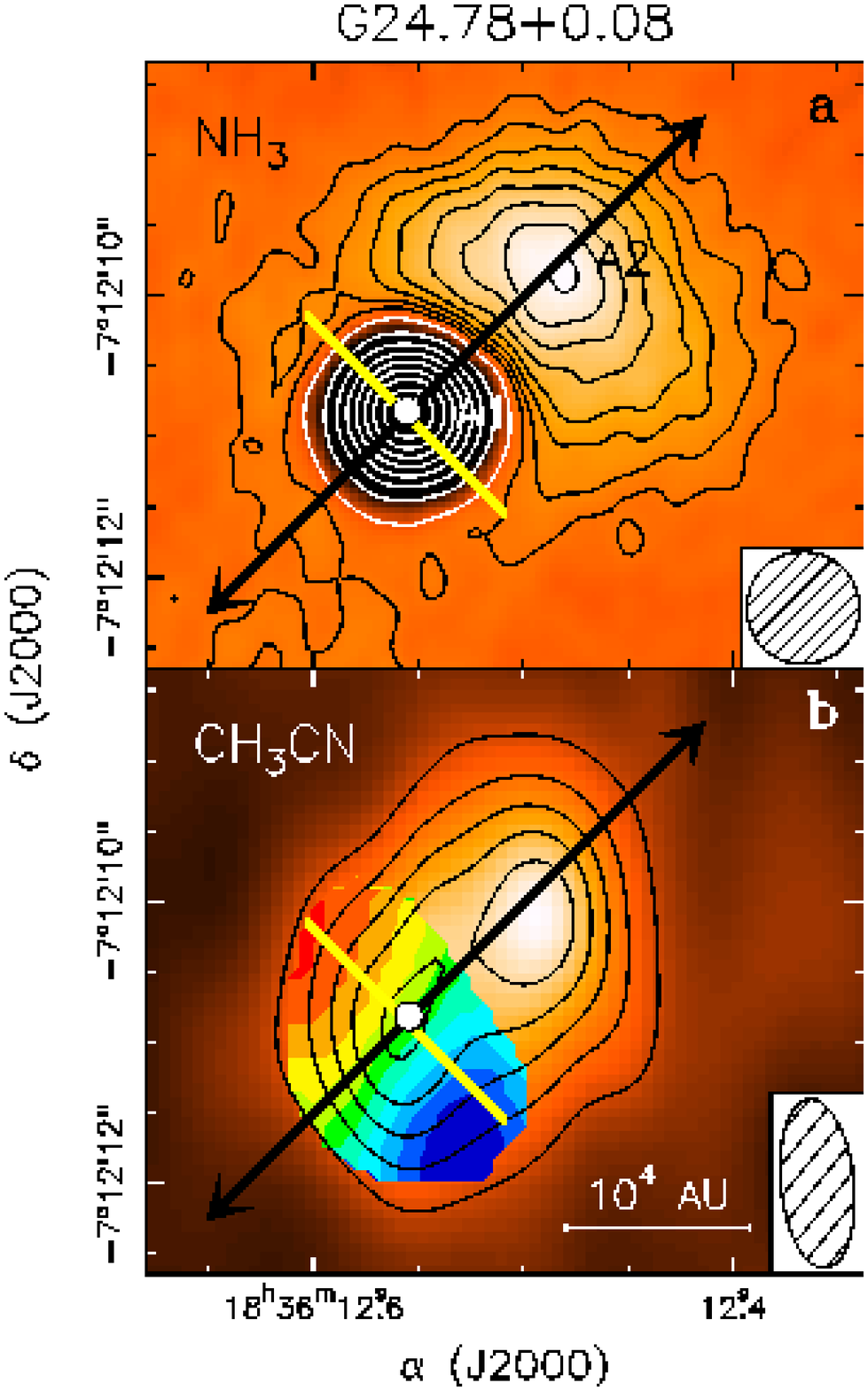}}}}
\caption{
}
\end{figure}

\begin{figure}
\centerline{\resizebox{8.9cm}{!}{\rotatebox{0}{\includegraphics{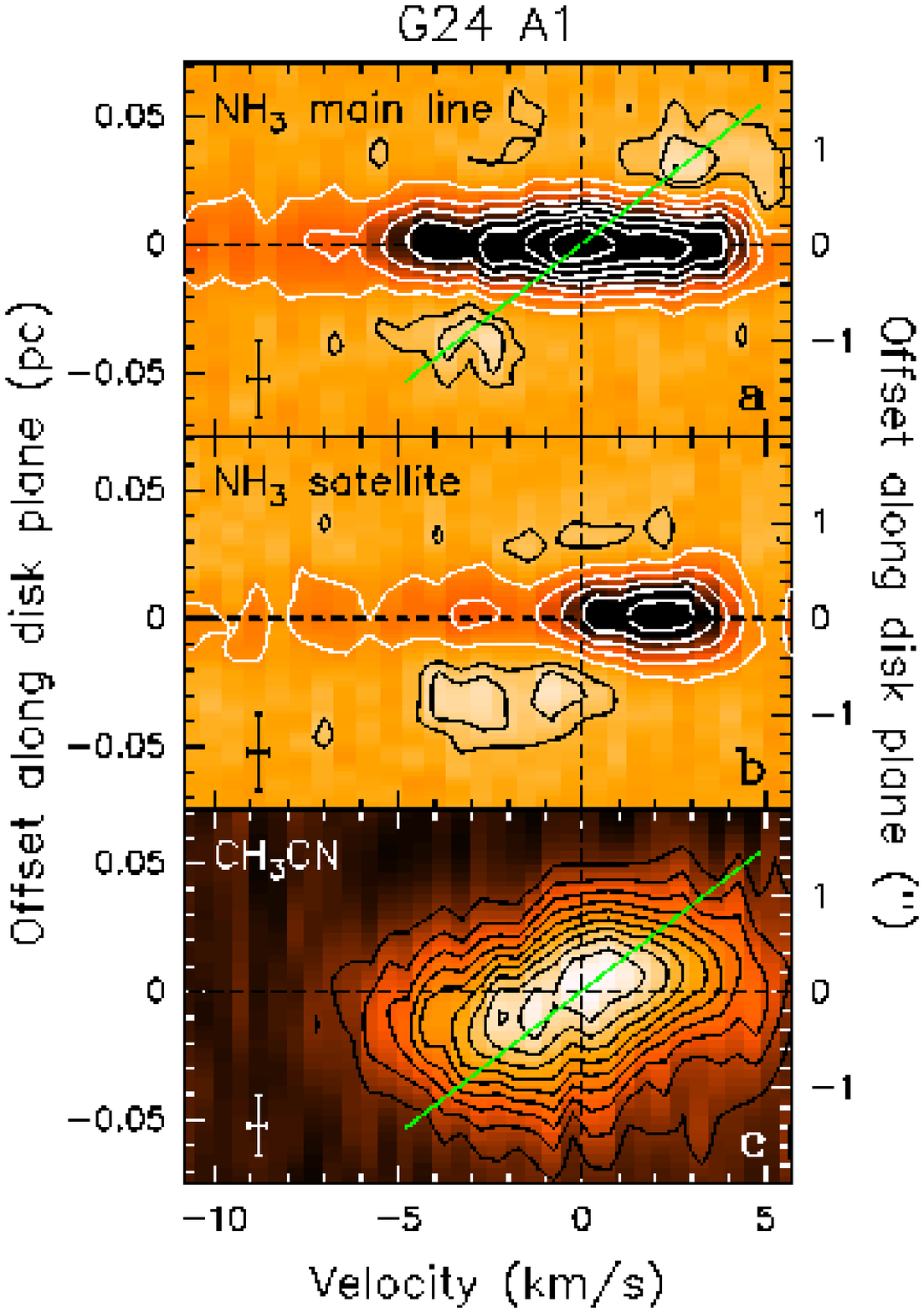}}}}
\caption{
}
\end{figure}

\end{document}